\documentclass[12pt]{article}
\usepackage{amsthm} 
\usepackage{amsmath} 
\usepackage{amsfonts}
\usepackage{amsxtra}
\usepackage{yhmath}
\usepackage{amssymb} 
\usepackage{mathrsfs} 
\usepackage{esint} 
\usepackage{empheq}
\usepackage{stmaryrd}
\usepackage{dsfont}
\usepackage{upgreek}
\usepackage[all]{xy}
\usepackage{graphicx}
\usepackage{latexsym}
\usepackage{color}
\usepackage{epsfig}
\usepackage{graphicx}
\usepackage{caption}
\usepackage{subcaption}
\usepackage{here}
\usepackage{float}
\usepackage{array}
\usepackage{tikz}
\usepackage{tikz-cd}
\usepackage{hyperref}
\usetikzlibrary{3d,arrows,calc,chains,matrix,positioning,scopes}

\DeclareSymbolFont{UPM}{U}{eur}{m}{n}
\DeclareMathSymbol{\uppartial}{0}{UPM}{"40}

\newcommand{\determinant}[2][]{\mathrm{det}_{#1}\mathopen{}\left(#2\right)}
\def\ba{\begin{array}}
	\def\ea{\end{array}}

\def\bea{\begin{eqnarray}}
	\def\eea{\end{eqnarray}}

\newcommand{\ensemblenombre}[1]{\mathbb{#1}}

\newcommand{\Z}{\ensemblenombre{Z}}

\newcommand{\slfrac}[2]{\left.#1\middle/#2\right.}

\newcommand{\abs}[1]{\left\lvert#1\right\rvert}

\theoremstyle{definition}

\theoremstyle{plain}

\begin{document}
	
	\pagestyle{empty}
	\setcounter{page}{0}
	\hspace{-1cm}

	\begin{center}
		{\Large {\bf Reshetikhin-Turaev construction and $\mathrm{U}(1)^n$ Chern-Simons partition function}}%
		\\[1.5cm]
		
		{\large M. Tagaris and F. Thuillier}
		\vspace{3mm}
		
		{\it Université Grenoble Alpes, USMB, CNRS, LAPTh, F-74000 Annecy, France, michail.tagaris@lapth.cnrs.fr and frank.thuillier@lapth.cnrs.fr}
	\end{center}
	
	\vskip 0.7 truecm

	\vspace{3cm}
	
	\centerline{{\bf Abstract}}
	In this article, we show that the $\mathrm{U}(1)^n$ Chern-Simons partition functions are related to Reshetikhin-Turaev invariants. In this abelian context, it turns out that the Reshetikhin-Turaev construction that yields these invariants relies on a ``twisted" category rather than a modular one. Furthermore, the Chern-Simons duality of the $\mathrm{U}(1)^n$ partition functions straightforwardly extend to the corresponding Reshetikhin-Turaev invariants.
	
	\vspace{2cm}
	
	\vfill
	
	\newpage
	\pagestyle{plain} \renewcommand{\thefootnote}{\arabic{footnote}}
	
	\section{Introduction}
	
	In a recent article \cite{KMTT25}, we studied a natural extension of the $\mathrm{U}(1)$ Chern-Simons theory on a closed $3$-manifold to a $\mathrm{U}(1)^n$ Chern-Simons theory. The partition function of such a theory has been determined thus providing a new family of manifold invariants. This also allowed us to see that the $\mathrm{U}(1)$ BF theory is nothing but a particular $\mathrm{U}(1)^2$ Chern-Simons theory. This is the equivalent of the Drinfeld center construction which allows to obtain the Turaev-Viro invariant from a Reshetikhin-Turaev construction \cite{MT17}. Furthermore, the use of the Deloup-Turaev reciprocity formula \cite{DT07} yields an expression of such a partition function that reminds us of a Reshetikhin-Turaev invariant \cite{RT91}. In this article, we show that indeed there is a Reshetikhin-Turaev construction which allows to associate to each $\mathrm{U}(1)^n$ Chern-Simons partition function a Reshetikhin-Turaev invariant. However, as we will see, this ``abelian" RT construction is based on a ``twisted" category and not on a modular one. We will explain where this difference with the non-abelian case comes from.

\vspace{0.5cm}

\textbf{Main result:} Any $\mathrm{U}(1)^n$ Chern-Simons partition function with coupling matrix $\mathbf{C}$ defined on a closed oriented smooth $3$-manifold $M$ coincides, up to some normalization factor, with an Reshetikhin-Turaev invariant whose construction relies on the quadratic form defined by the symmetrized coupling matrix $\mathbf{K} = \mathbf{C} + \mathbf{C}^\dagger$ and on the linking matrix $\mathbf{L}$ of an integer surgery of $M$ in $S^3$. Moreover, if the surgery is even then this Reshetikhin-Turaev invariant can in turn be seen as the partition function of a $\mathrm{U}(1)^m$ Chern-Simons with symmetrized coupling matrix $\mathbf{L}$ defined on a manifold with linking matrix $\mathbf{K}$. This defines the Chern-Simons duality which straightforwardly extends to a Reshetikhin-Turaev duality between abelian Reshetikhin-Turaev invariants.

\vspace{0.5cm}
	
	In a first section we will recall shortly how the $\mathrm{U}(1)^n$ Chern-Simons action is defined in the Deligne-Beilinson cohomology \cite{B93,BGST05,GT08,GT13,GT14} context and how the partition function is obtained. Eventually, a reciprocity procedure will be applied to this partition function thus yielding an expression that recalls the one of a Reshetikhin-Turaev invariant.
	
	In a second section we will explain why the partition function, as well as its reciprocal expression, i.e., its expression after a reciprocity procedure has been applied, defines a manifold invariant. Then, we will present a construction inspired from the Reshetikhin-Turaev one which yields the reciprocal expression of the partition function. Finally, we will discuss Chern-Simons duality in this context. Let us point out that the Reshetikhin-Turaev invariant we obtain from a $\mathrm{U}(1)^n$ Chern-Simons partition function can also be seen as a generalization of closed oriented 3-manifold invariant introduced by Murakami et al. in \cite{MOO92}.

	Although all along this article $M$ is a closed oriented smooth $3$-dimensional manifold, the construction and all the results straightforwardly extend to higher dimensional closed oriented smooth manifolds (of appropriate dimension).

    \section{Abelian \texorpdfstring{$\mathrm{U}(1)^n$}{U(1)	extasciicircum n} Chern-Simons theory}
	
	Let us recall that the action of the usual $\mathrm{U}(1)$ Chern-Simons theory is the $\mathbb{R}/\mathbb{Z}$-valued functional defined as
    \begin{equation}
    \label{CSaction}
	S[A] = \oint_M A \star A \, ,
    \end{equation}
    where $A$ denotes an element of $H^1_D(M)$, the group of gauge classed of $\mathrm{U}(1)$ gauge fields on $M$. This group can be canonically identified with the first Deligne-Beilinson cohomology group of $M$. The DB product $\star$ is a commutative operation on $H^1_D(M)$ which takes values in the quotient space $\Omega^3(M)/\Omega^3_\mathbb{Z}(M)$ where $\Omega^3_\mathbb{Z}(M)$ is the space of closed $3$-forms with integral periods on $M$, the integral which defines the action (\ref{CSaction}) thus taking its values in $\mathbb{R}/\mathbb{Z}$ \cite{B93}. A natural generalization of the $\mathrm{U}(1)$ Chen-Simons theory is obtained by considering a collection of $n$ quantum fields $A_i$ together with a collection $C_{ij}$ of integers such that \cite{KMTT25}
    \begin{equation}
    \label{multiCSaction}
	S_\mathbf{C}[\mathbf{A}] = \sum_{i,j} C_{ij} \oint_M A_i \star A_j := \oint_M \mathbf{A} \star_\mathbf{C} \mathbf{A} \, ,
    \end{equation}
	where $\mathbf{A}$ denotes the column whose entries are the quantum fields $A_i$. The coupling constants $C_{ij}$ are necessarily integers since the action is defined modulo integers. The collection of integers $C_{ij}$ defines a $n \times n$ matrix $\mathbf{C}$ which is referred to as the \textbf{coupling matrix} of the $\mathrm{U}(1)^n$ Chern-Simons theory with action (\ref{multiCSaction}). In fact, since the DB product on $H^1_D(M)$ is commutative, it is always possible to turn $\mathbf{C}$ into an upper triangular matrix, a convention that we adopt from now on.
    
	The partition function associated to the action (\ref{multiCSaction}) is defined as the formal functional integral
    \begin{equation}
	\mathcal{Z}_\mathbf{C} = \mathcal{N} \int D \! \mathbf{A} \, e^{2 i \pi S_\mathbf{C}[\mathbf{A}]} \, .
    \end{equation}
The normalization factor $\mathcal{N}$ is chosen in such way that an ``irrelevant" infinite dimensional contribution to the functional integration defining $\mathcal{Z}_\mathbf{C}$ is eliminated \cite{KMTT25}. After some simple computations, the remaining finite contributions yield
	\begin{equation}
    \label{partU1n}
		\mathcal{Z}_\mathbf{C}
		= \sum_{\mathbf{X} \in \left(TH^2(M)\right)^n}
		e^{-2 i \pi \left(\mathbf{C} \otimes \mathbf{Q}\right) (\mathbf{X},\mathbf{X}) } \, ,
	\end{equation}
where $\mathbf{X}$ is a column matrix whose entries $\mathbf{x}_i$ are elements of $TH^{2}(M)$ and $\mathbf{Q}: TH^{2}(M) \times TH^{2}(M) \rightarrow \mathbb{R}/\mathbb{Z}$ is the linking form so that
    \begin{equation}
    \label{evalCQ}
        \left(\mathbf{C} \otimes \mathbf{Q}\right) (\mathbf{X},\mathbf{X}) = \sum_{i,j} C_{i,j} \, \mathbf{Q}(\mathbf{x}_i,\mathbf{x}_j) \, .
    \end{equation}
The expression (\ref{partU1n}) is considered as the partition function of the $\mathrm{U}(1)^n$ Chern-Simons theory with coupling matrix $\mathbf{C}$.
    
It is possible to express $\mathcal{Z}_\mathbf{C}$ in a slightly different way. To that end, let us introduce the following even symmetric matrix
	\begin{equation}
		\mathbf{K} = \mathbf{C} + \mathbf{C}^\dagger \, .
	\end{equation}
We recall that a symmetric integer matrix is even if its diagonal entries are even. We also assume that $\mathbf{K}$ is non-degenerate. Then, if the same representatives of the $\mathbf{x}_i$ are used in the evaluation of the right-hand side of (\ref{evalCQ}), the partition function can be written as
	\begin{equation}
    \label{partKQ}
		\mathcal{Z}_\mathbf{C}
		= \sum_{\mathbf{X} \in \left(TH^{2}(M)\right)^n}
		e^{- i \pi \left(\mathbf{K}\otimes\mathbf{Q}\right) (\mathbf{X}) } \, ,
	\end{equation}
with $\left(\mathbf{K} \otimes \mathbf{Q}\right) (\mathbf{X}) = \sum_{i,j} K_{i,j} \, \mathbf{Q}(\mathbf{x}_i,\mathbf{x}_j)$. Let us recall that $\mathbf{Q}$ is first defined on the representatives of the elements of $TH^{2}(M)$. Now, if $\mathbf{K}$ is \underline{even} symmetric and if the \underline{same} representative of $\mathbf{x}_i$ is used in the evaluation of $\mathbf{Q}(\mathbf{x}_i,\mathbf{x}_i)$, then the quantity $\sum_{i,j} K_{i,j} \, \mathbf{Q}(\mathbf{x}_i,\mathbf{x}_j)$ is defined modulo $2$. In other words, when writing $\left(\mathbf{K} \otimes \mathbf{Q}\right) (\mathbf{X})$ we see $(\mathbf{K} \otimes \mathbf{Q})$ as a quadratic form on the finite group $\left(TH^{2}(M)\right)^n$ \cite{W64,W72,DT07}. From now on, we adopt this notation convention.
    
A reciprocity procedure can be applied to the above expression of the partition function. This is a generalization of the usual reciprocity procedure to tensor product as explained in \cite{DT07}. Let us note that the (matrix of the) linking form $\mathbf{Q}$ can always be seen as the inverse of some $m \times m$ symmetric, even and non-degenerate matrix $\mathbf{L}$, so that $TH^2(M) \simeq \mathbb{Z}^m/\mathbf{L}\mathbb{Z}^m$. Of course, in the case of a closed oriented smooth $3$-manifold $M$, the matrix $\mathbf{L}$ can be obtained as the linking matrix of an even surgery of $M$ in $S^3$ \cite{S12}. The non-degeneracy of the corresponding linking matrix implies that $H^2(M)$ is finite, i.e., it has no free sector. Let us stress out that although surgery may be useful, the computation of the partition function does not rely on it. The reciprocity procedure yields the following relation
	\begin{align}
    \label{Recipro}
		\nonumber
		& \sum_{ \mathbf{X} \in\left(\slfrac{\Z^{m}}{\mathbf{L}\Z^{m}}\right)^{n}}
	    e^{-i \pi \left(\mathbf{K}\otimes\mathbf{L}^{-1}\right) (\mathbf{X})}  \\ 
        &= \frac{\abs{\determinant{\mathbf{L}}}^{\frac{n}{2}}}{\abs{\determinant{\mathbf{K}}}^{\frac{m}{2}}} e^{- \frac{i\pi}{4}\sigma\left(\mathbf{K}\right)\sigma\left(\mathbf{L}\right)}
		\sum\limits_{\mathbf{U} \in \left(\slfrac{\Z^{n}}{\mathbf{K}\Z^{n}}\right)^{m}}
		e^{i \pi \left(\mathbf{L}\otimes\mathbf{K}^{-1}\right) (\mathbf{U})}
	\end{align}
    On both sides of this relation the notation convention introduced after (\ref{partKQ}) is used. The sum on the right-hand side is an invariant of $M$ in the sense that it is invariant under the Kirby moves applied to $\mathbf{L}$ \cite{T23} . Let us point out that the partition function $\mathcal{Z}_\mathbf{C}$ relies on $(TH^2(M),\mathbf{Q})$, and since many non-diffeomorphic closed $3$-manifolds share these data, this invariant is ``weakly" classifying.
    
    As suggested by the above reciprocity formula, the invariant defined by $\mathcal{Z}_\mathbf{C}$ should be related to some Reshetikhin-Turaev invariant, just as in the standard $\mathrm{U}(1)$ Chern-Simons theory. The second section of this article is dedicated to this question. Furthermore, since the Reshetikhin-Turaev invariant construction relies on integer surgery of $M$, the above reciprocity formula is nothing but a surgery formula and as such it should extend to expectation values of Wilson observables.

    \section{Reshetikhin-Turaev invariant}
	
	The general Reshetikhin-Turaev construction for invariants of closed oriented smooth $3$-manifolds is based on modular category theory \cite{T10}. However, in the abelian case, i.e., in the case which is related to the $\mathrm{U}(1)$ Chern-Simons partition function, the category on which this invariant can be obtained (and thus, up to reciprocity, coincides with the partition function $\mathrm{U}(1)$ Chern-Simons partition function) is not modular \cite{MT16}. In fact, this non-modularity is the reason why in the abelian case the square modulus of the Reshetikhin-Turaev invariant does not coincide with the Turaev-Viro invariant, unlike what happens in the $SU(N)$ case. Let us point out that the case where $G_\mathbf{K} = \mathbb{Z}/N\mathbb{Z}$ has been studied in detail in \cite{MOO92}.
	
	Let $\mathbf{K}$ be an even symmetric non-degenerate $n \times n$ integer matrix. We also denote by $\mathbf{K}:\mathbb{Z}^n \rightarrow \mathbb{Z}^n$ the homomorphism that this matrix defines. This homomorphism defines in turn a bihomomorphism $\hat{\mathbf{K}} : \mathbb{Z}^n \times \mathbb{Z}^n \rightarrow \mathbb{Z}$ according to
    \begin{equation}
        \hat{\mathbf{K}}(\vec{u},\vec{v}) = <\vec{u},\mathbf{K}\vec{v}> \, ,
    \end{equation}
    where $<\, \;,\,\,>: \mathbb{Z}^n \times \mathbb{Z}^n \rightarrow \mathbb{Z}$ is the canonical pairing. Then, as $\mathbf{K}$ is non-degenerate, the quotient $G_{\mathbf{K}} = \mathbb{Z}^n/\mathbf{K}\mathbb{Z}^n$ is a finite abelian group such that if $\mathbf{u} \in G_{\mathbf{K}}$ denotes the equivalence class of $\vec{u} \in \mathbb{Z}^n$, then for any $\vec{w} \in \mathbb{Z}^n$ the equivalence class of $\mathbf{K} \vec{w}$ is $\mathbf{0} \in G_{\mathbf{K}}$. The matrix $\mathbf{K}^{-1}$ generates a linking form $\mathbf{Q}_\mathbf{K}$ on $G_{\mathbf{K}}$, i.e., an $\mathbb{R}/\mathbb{Z}$-valued non-degenerate symmetric bihomomorphism on $G_{\mathbf{K}}$ according to 
    \begin{equation}
    \label{DefQ_K}
        \mathbf{Q}_\mathbf{K}(\mathbf{u},\mathbf{v}) = < \vec{u}, \mathbf{K}^{-1} \vec{v}>_\mathbb{Q} \, ,
    \end{equation}
    where $\vec{u}$ (resp. $\vec{v}$) is a representative of $\mathbf{u}$ (resp. $\mathbf{v}$) and $<\, \;,\,\,>_\mathbb{Q}$ is the rational extension of $<\, \;,\,\,>$. This definition is meaningful since
    \begin{equation}
        <\vec{u} + \mathbf{K}\vec{w}_1 , \mathbf{K}^{-1} (\vec{v} + \mathbf{K}\vec{w}_2)>_\mathbb{Q} = <\vec{u},\mathbf{K}^{-1}\vec{v}>_\mathbb{Q}  \text{ mod } \mathbb{Z} \, ,
    \end{equation}
     for any $\vec{w}_1$ and $\vec{w}_2$. A one dimensional representation of $G_{\mathbf{K}}$ is a group homomorphism $R: G_{\mathbf{K}} \rightarrow Gl(\mathbb{C})$. The set of irreducible one-dimensional representations of $G_{\mathbf{K}}$ is generated by 
    \begin{equation}
        R_{\mathbf{u}}: \mathbf{v} \mapsto e^{2 i \pi  \mathbf{Q}_\mathbf{K} (\mathbf{u} , \mathbf{v}) }  \, ,
    \end{equation}
    with $\mathbf{u} \in G_\mathbf{K}$. The irreducible one-dimensional representations of $G_{\mathbf{K}}$ can be seen as the objects of a category $\mathbb{C}^{G_{\mathbf{K}}}$, the morphisms of which are the so-called natural transformations. 
    
    The category $\mathbb{C}^{G_{\mathbf{K}}}$ can be endowed with a natural tensor product by setting:
    \begin{equation}
        R_{\mathbf{u}_1} \otimes R_{\mathbf{u}_2} := R_{\mathbf{u}_1 + \mathbf{u}_2} \, .
    \end{equation}
    Obviously, $R_{\mathbf{u}_1} \otimes R_{\mathbf{u}_2} = R_{\mathbf{u}_2} \otimes R_{\mathbf{u}_1}$, the identity object of $\mathbb{C}^{G_{\mathbf{K}}}$ is the trivial representation $R_{\mathbf{0}}$, and the corresponding associator is trivial. This makes $\mathbb{C}^{G_{\mathbf{K}}}$ a symmetric strict monoidal category.
    
    A duality is defined on $\mathbb{C}^{G_{\mathbf{K}}}$ by setting:
    \begin{equation}
        R_{\mathbf{u}}^\star: \mathbf{v} \mapsto e^{- 2 i \pi  \mathbf{Q}_K (\mathbf{u} , \mathbf{v}) } = \overline{R_{\mathbf{u}}(\mathbf{v})} \, .
    \end{equation}
    This eventually turns $\mathbb{C}^{G_{\mathbf{K}}}$ into a rigid symmetric strict monoidal category.
    
    The dimension of an object $R_{\mathbf{u}}$ of $\mathbb{C}^{G_{\mathbf{K}}}$ is defined to be
    \begin{equation}
       \text{dim} (R_{\mathbf{u}}) = Tr(\mathbb{I}d) = 1 \, ,
    \end{equation}
    where $\mathbb{I}d: Gl(\mathbb{C}) \rightarrow Gl(\mathbb{C})$ is the identity automorphism. The global dimension $\mathcal{D}_{\mathbf{K}}$ of $\mathbb{C}^{G_{\mathbf{K}}}$ is then
    \begin{equation}
       (\mathcal{D}_{\mathbf{K}})^2 = \sum_{\mathbf{u} \in G_{\mathbf{K}}} (\text{dim} (R_{\mathbf{u}}))^2 = | G_{\mathbf{K}} | \, ,
    \end{equation}
    where $| G_{\mathbf{K}} |$ denotes the order of the group $G_{\mathbf{K}}$. Hence
    \begin{equation}
       \mathcal{D}_{\mathbf{K}} = \sqrt{|G_{\mathbf{K}}|} \, ,
    \end{equation}
    
    Now, we provide $\mathbb{C}^{G_{\mathbf{K}}}$ with a twist by introducing the collection of morphisms $\theta_{\mathbf{u}}: R_{\mathbf{u}} \rightarrow R_{\mathbf{u}}$ defined by:
    \begin{equation}
    \label{twist}
        \boldsymbol{\theta}_{\mathbf{u}} = e^{i \pi \mathbf{Q}_\mathbf{K} (\mathbf{u})} 
        \, ,
    \end{equation}    
    with $\mathbf{Q}_\mathbf{K} (\mathbf{u}) = \mathbf{Q}_\mathbf{K} (\mathbf{u} , \mathbf{u})$, the notation convention introduced after (\ref{partKQ}) thus being used. Endowed with the above twist, we say that $\mathbb{C}^{G_{\mathbf{K}}}$ is a twisted category. This is the only structure that we will need in the abelian Reshetikhin-Turaev construction. 

    The Reshetikhin-Turaev construction requires to provide $\mathbb{C}^{G_{\mathbf{K}}}$ with an $S$-matrix. This is achieved by setting
    \begin{equation}
    \label{Smatdef}
        \mathbf{S}_{\mathbf{u} ,\mathbf{v}} = \boldsymbol{\theta}_{\mathbf{u} + \mathbf{v} } \boldsymbol{\theta}_{\mathbf{u}}^{-1} \boldsymbol{\theta}_{\mathbf{v}}^{-1} = e^{2 i \pi \mathbf{Q}_\mathbf{K} (\mathbf{u} , \mathbf{v})}  \, ,
    \end{equation} 
    so that, if $(\mathbf{u}_i)_{i=1,\cdots,m}$ is a collection of elements of ${G_{\mathbf{K}}}$, we have
    \begin{equation}
    \label{Smatrix}
        \mathbf{S}_{\mathbf{u}_i,\mathbf{u}_j} = e^{2 i \pi \mathbf{Q}_\mathbf{K} (\mathbf{u}_i , \mathbf{u}_j)} 
        \, .
    \end{equation}

\vspace{0.5cm}
\noindent \textbf{Remark:} In the standard Reshetikhin-Turaev procedure, we should have first introduced a braiding on $\mathbb{C}^{G_{\mathbf{K}}}$ and then a twist which is compatible with this braiding. Although it exists, such a braiding is not essential in our abelian context. Accordingly, we do not care if $\mathbf{S}_{\mathbf{u} ,\mathbf{v}} = \mathbf{c}_{\mathbf{u} ,\mathbf{v}} \mathbf{c}_{\mathbf{v} ,\mathbf{u}}$ for some braiding $\mathbf{c}$ on $\mathbb{C}^{G_{\mathbf{K}}}$. We do not even need $\mathbf{S}$ to be non-degenerate. In fact, in the usual $\mathrm{U}(1)$ Chern-Simons theory, the $S$-matrix used to construct the associated Reshetikhin-Turaev invariant is degenerate, while in the $\mathrm{U}(1)$ BF case it is not \cite{MT16}.

\vspace{0.5cm}

Still following the Reshetikhin-Turaev procedure, we introduce the quantity
    \begin{equation}
        \Delta_{\mathbf{K}} = \sum_{\mathbf{u} \in G_{\mathbf{K}}} (\theta_{\mathbf{u}})^{-1} (\text{dim} (R_{\mathbf{u}}))^2 = \sum_{\mathbf{u} \in G_{\mathbf{K}}} e^{- i \pi \mathbf{Q}_\mathbf{K} (\mathbf{u})} \, .
    \end{equation} 
    Finally, let $\mathbf{L}$ be some even linking $m \times m$ matrix (as the one generated by an even integer surgery in $S^3$ of a closed oriented $3$-manifold $M$). Then, the Reshetikhin-Turaev invariant we are looking for is
    \begin{equation}
    \label{RTinv}
        \mathcal{RT}_{\mathbf{K}}(\mathbf{L}) = (\Delta_{\mathbf{K}})^{\sigma(\mathbf{L})} (\mathcal{D}_{\mathbf{K}})^{-\sigma(\mathbf{L}) -m} \sum_{\mathbf{u}_1 \in G_{\mathbf{K}}} \cdots \sum_{\mathbf{u}_m \in G_{\mathbf{K}}} \prod\limits_{i} (\theta_{\mathbf{u}_i})^{L_{ii}} \prod\limits_{k < l} (\mathbf{S}_{\mathbf{u}_k,\mathbf{u}_l})^{L_{kl}} 
        \, .
    \end{equation} 
    Let us first deal with the factor in front of the sums in the right-hand side of the above expression. From \cite{W72} we deduce that $\Delta_{\mathbf{K}} = e^{- \frac{i \pi }{4} \sigma(\mathbf{K})} \sqrt{|G_{\mathbf{K}}|} = e^{- \frac{i \pi }{4} \sigma(\mathbf{K})} \mathcal{D}_{\mathbf{K}}$. Thus, the front factor in the right-hand side of (\ref{RTinv}) is $e^{- \frac{i \pi }{4} \sigma(\mathbf{K})\sigma(\mathbf{L})} |G_{\mathbf{K}}|^{-\frac{m}{2}}$. Now, since $|G_{\mathbf{K}}| = |\determinant{\mathbf{K}}|$, we conclude that this front factor coincides with the front factor in the right-hand side of (\ref{Recipro}). 

    Concerning the sums in the right-hand side of (\ref{RTinv}), by injecting the explicit the expression of the twist and the one of the $S$-matrix, respectively defined by (\ref{twist}) and (\ref{Smatrix}), and by using the fact that $\mathbf{L}$ is even and symmetric, we deduce that
    \begin{equation}
        \sum_{\mathbf{u}_1 \in G_{\mathbf{K}}} \cdots \sum_{\mathbf{u}_m \in G_{\mathbf{K}}} e^{ i \pi L_{ij} \mathbf{Q}_K (\mathbf{u}_i , \mathbf{u}_j)} = \sum_{\mathbf{U} \in (G_{\mathbf{K}})^m} e^{ i \pi \left(\mathbf{L} \otimes \mathbf{Q}_\mathbf{K} \right) (\mathbf{U}) } \, ,
    \end{equation}
where $\mathbf{U}$ denotes the collection of the $\mathbf{u}_i \in G_{\mathbf{K}}$. Hence, we obtain
    \begin{equation}
    \label{RTinvbis}
        \mathcal{RT}_{\mathbf{K}}(\mathbf{L}) = e^{- \frac{i \pi }{4} \sigma(\mathbf{K})\sigma(\mathbf{L})} |\determinant{\mathbf{K}}|^{-\frac{m}{2}} \sum_{\mathbf{U} \in (G_{\mathbf{K}})^m} e^{ i \pi \left(\mathbf{L} \otimes \mathbf{Q}_\mathbf{K} \right) (\mathbf{U}) }
        \, .
    \end{equation}
    By combining this with on the one hand the expression of the Chern-Simons partition function and on the other hand the reciprocity formula, we finally obtain
    \begin{equation}
    \label{result}
        \mathcal{Z}_\mathbf{C} = \abs{\determinant{\mathbf{L}}}^{\frac{n}{2}} \mathcal{RT}_{\mathbf{K}}(\mathbf{L}) = \abs{\determinant{\mathbf{K}}}^{\frac{m}{2}} \mathcal{RT}_{\mathbf{L}}(\mathbf{K}) \, .
    \end{equation}
    As a particular case, we recover the standard $\mathrm{U}(1)$ case.

\vspace{0.5cm}
 
\noindent \textbf{Remarks}:
    
    1) As already said, no braiding is explicitly required in the Reshetikhin-Turaev construction we have proposed. Nonetheless, there is a braiding associated with a twist as defined by (\ref{twist}). To see this, let us use the standard decomposition
    \begin{equation}
        G_{\mathbf{K}} \simeq \mathbb{Z}_{N_1} \times \cdots \times \mathbb{Z}_{N_r} \, ,
    \end{equation}
    with $N_i$ dividing $N_{i+1}$. If $\mathbf{g}_i$, $i=1, \cdots, r$, is a set of generators of $G_{\mathbf{K}}$, then any $\mathbf{u} \in G_{\mathbf{K}}$ can be written as $\sum_i u_i \mathbf{g}_i$. Then, according to \cite{GLO18} the expression
    \begin{equation}
    \label{braiding}
        \mathbf{c}_{\mathbf{u},\mathbf{v}} =  \prod\limits_{i} (\boldsymbol{\theta}_{\mathbf{g}_i})^{u_i v_i} \prod\limits_{k < l} (\mathbf{S}_{\mathbf{g}_k,\mathbf{g}_l})^{u_k v_l} \, ,
    \end{equation}
defines a braiding on $\mathbb{C}^{G_{\mathbf{K}}}$. Note that in the above relation the $u_i$ are not the components of $\Vec{U}$ in the canonical basis of $\mathbb{Z}^n$ but components in the basis of $G_{\mathbf{K}}$ defined by generators $\mathbf{g}_i$. The reason why, in this abelian context, the braiding is defined by the twist is twofold. Firstly, if $\mathbf{K}$ is a non-degenerate symmetric even integer $n \times n$ matrix then it canonically generates a quadratic form $\mathbf{Q}_\mathbf{K}$ on the finite abelian group $\mathbf{G}_\mathbf{K}=\mathbb{Z}^n/\mathbf{K}\mathbb{Z}^n$. Secondly, there is an isomorphism between the set of quadratic forms on $\mathbf{G}_\mathbf{K}$ and the abelian cohomology group of $\mathbf{G}_\mathbf{K}$, any class of which is associated with a collection of (equivalent) braidings \cite{GLO18}.
    
2) As we don't require the $\mathbf{S}$ matrix to be non-degenerate, we don't need $\mathbb{C}^{G_\mathbf{K}}$ to be a modular category. For instance, in the case of the standard $\mathrm{U}(1)$ Chern-Simons theory the $S$-matrix is degenerate.

\vspace{0.5cm}

\textbf{Summary:} Let $M$ be a closed oriented smooth $3$-manifold with linking form $\mathbf{Q}_M: T_1(M) \times T_1(M) \rightarrow \mathbb{Q}/\mathbb{Z}$. Let $\mathbf{L}$ be an $n \times n$ linking matrix of $M$, i.e., the linking matrix of some integer surgery of $M$ in $S^3$, and let $\mathbf{C}$ be an $m \times m$ integer matrix. The partition function of the $\mathrm{U}(1)^n$ Chern-Simons theory with coupling matrix $\mathbf{C}$ on $M$ only depends on the $m \times m$ even symmetric integer matrix $\mathbf{K} = \mathbf{C} + \mathbf{C}^\dagger$ and on the linking form $\mathbf{Q}_M$. By applying the Deloup-Turaev reciprocity procedure on such a partition function one obtains a quantity which turns out to be a Reshetikhin-Turaev invariant built with the help of the quadratic form canonically defined by $\mathbf{K}$ and the linking matrix $\mathbf{L}$. Now, if the linking matrix $\mathbf{L}$ is chosen to be even, which is always possible, then the Reshetikhin-Turaev invariant thus obtained can be seen (without any reciprocity procedure) as the partition function of a $\mathrm{U}(1)^m$ Chern-Simons theory with coupling matrix $\tilde{\mathbf{C}}$ such that $\mathbf{L} = \tilde{\mathbf{C}} + \tilde{\mathbf{C}}^\dagger$ and linking form $\mathbf{Q}_\mathbf{K}$. This theory is said to be dual to the original one. By performing a new reciprocity procedure on this partition function, one obtains a quantity which is again an Reshetikhin-Turaev invariant, dual from the original one. This can be represented as follows:
\[\begin{gathered}
  {\mathcal{Z}_{K,{Q_M}}} \xleftrightarrow{{CS{\text{ duality}}}}{\mathcal{Z}_{L,{Q_K}}} \hfill \\
  \quad { \updownarrow _{recip.}} \quad \quad \quad \quad \quad { \updownarrow _{recip.}} \hfill \\
  R{T_{{Q_K},L}} \xleftrightarrow{{RT{\text{ duality}}}}R{T_{{Q_L},K}} \hfill \\ 
\end{gathered} \]
The vertical left (resp. right) arrow holds true if $\mathbf{K}$ (resp. $\mathbf{L}$) is  even while the horizontal arrows hold true if both $\mathbf{K}$ and $\mathbf{L}$ are even.

	\section{Some examples}
	
	1) The standard $\mathrm{U}(1)$ Chern-Simons theory: in this case the coupling matrix is the $1 \times 1$ matrix $\mathbf{C} = (k)$, the associated $\mathbf{K}$ matrix thus being $\mathbf{K} = (2k)$. It is quite straightforward to check that the previous procedure yields the standard Reshetikhin-Turaev invariant based on the finite group $\mathbb{Z}_{2k}$. Note that in this very simple case $\mathbb{Z}/\mathbf{K}\mathbb{Z} = \mathbb{Z}_{2k}$, as expected \cite{GT08,GT13,GT14}.

    \noindent 2) The standard $\mathrm{U}(1)$ BF theory \cite{MT16,MT16B,MT17}: in this case the coupling matrix is
    \[ \mathbf{C} = \left( {\begin{array}{*{20}{c}}
    0 & k\\
    0 & 0
    \end{array}} \right) \, . \]
	The associated $K$ matrix is then
    \[ \mathbf{K} = \left( {\begin{array}{*{20}{c}}
    0 & k\\
    k & 0
    \end{array}} \right) \, , \]
    In this still simple case, $\mathbb{Z}^2/\mathbf{K}\mathbb{Z}^2 = \mathbb{Z}_{k} \times \mathbb{Z}_{k}$ is the Drinfeld Center of $\mathbb{Z}_{k}$ and the Reshitikhin-Turaev procedure yields a Turaev-Viro invariant as explained in \cite{MT16B}.

    \noindent 3) A less trivial example is provided by the coupling matrix
    \[ \mathbf{C} = \left( {\begin{array}{*{20}{c}}
    k & k\\
    0 & k
    \end{array}} \right) \, , \]
    the $K$ matrix of which is
    \[ \mathbf{K} = \left( {\begin{array}{*{20}{c}}
    2k & k\\
    k & 2k
    \end{array}} \right) \, , \]
    with $\determinant{\mathbf{K}} = 3 k^2$. It is quite obvious that $\mathbb{Z}^2/\mathbf{K}\mathbb{Z}^2 = \mathbb{Z}_{k} \times \mathbb{Z}_{3k}$. A set of generators for this finite group is provided by
    \[\mathbf{g}_1 = \left( {\begin{array}{*{20}{c}}
    2 \\
    1
    \end{array}} \right)\quad ,\quad \mathbf{g}_{2} = \left( {\begin{array}{*{20}{c}}
    1 \\
    1
    \end{array}} \right) \, . \]
    The inverse of $\mathbf{K}$, 
    \[ \mathbf{Q}_\mathbf{K} = \left( {\begin{array}{*{20}{c}}
    \frac{2}{3k} & -\frac{1}{3k} \\
    -\frac{1}{3k} & \frac{2}{3k}
    \end{array}} \right) \, , \]
    defines our linking form on $\mathbb{Z}_{k} \times \mathbb{Z}_{3k}$. In the basis $(\mathbf{g}_1,\mathbf{g}_2)$ of $\mathbb{Z}^2$, the matrix of the linking form $\mathbf{Q}_\mathbf{K}$ is
    \[ \mathbf{Q}_\mathbf{K} = \left( {\begin{array}{*{20}{c}}
    \frac{2}{k} & \frac{1}{k} \\
    \frac{1}{k} & \frac{2}{3k}
    \end{array}} \right) \, . \]
    
    According to definition (\ref{twist}), the associated twist reads
    \begin{equation}
        \mathbf{\theta}_{\mathbf{u}} = e^{2 i \pi (\frac{u_1^2}{k} + \frac{u_1 u_2}{k} + \frac{u_2^2}{3k})} \, ,
    \end{equation}
    with $\mathbf{u} = u_1 \mathbf{g}_1 + u_2 \mathbf{g}_2$. This twist is well-defined because $u_1$ is $k$-periodic and $u_2$ is $3k$-periodicity. Note that the above expression of $\mathbf{\theta}_{\mathbf{u}}$ is automatically well-defined if we take $u_1 \in \left\{0 ,  \cdots , k-1  \right\}$ and $u_2 \in \left\{0 ,  \cdots , 2k-1  \right\}$ as this ensures the requirement that the same representative of $\mathbf{u}$ must be used in the evaluation of $\mathbf{\theta}_{\mathbf{u}}$.
    The expression of the $S$-matrix, as defined in (\ref{Smatrix}), is
    \begin{equation}
        \mathbf{S}_{\Vec{U},\Vec{V}} = e^{2 i \pi (\frac{2 u_1 v_1}{k} + \frac{u_1 v_2 + u_2 v_1}{k} + \frac{2 u_2 v_2}{3k})} \, .
    \end{equation}
    with $\mathbf{u} = u_1 \mathbf{g}_1 + u_2 \mathbf{g}_2$ and $\mathbf{v} = v_1 \mathbf{g}_1 + v_2 \mathbf{g}_2$. Once more, the periodicity of the components of these vectors ensures that the $S$-matrix is well-defined. The (inessential) braiding, as defined by (\ref{braiding}), takes the form
    \begin{equation}
        \mathbf{c}_{\mathbf{u},\mathbf{v}} = (\mathbf{\theta}_{\mathbf{g}_1})^{u_1 v_1} (\mathbf{\theta}_{\mathbf{g}_2})^{u_2 v_2} (\mathbf{S}_{\mathbf{g}_1,\mathbf{g}_2})^{u_1 v_2} = e^{2 i \pi (\frac{u_1 v_1}{k} + \frac{u_1 v_2}{k} + \frac{u_2 v_2}{3k})} \, ,
    \end{equation}
    which is again meaningful thanks to the periodicity of $u_1$,$v_1$, $u_2$ and $v_2$. Note that
    \begin{equation}
        \mathbf{\theta}_{\mathbf{u} + \mathbf{v}} \mathbf{\theta}_{\mathbf{u}}^{-1} \mathbf{\theta}_{\mathbf{v}}^{-1} = \mathbf{S}_{\mathbf{u},\mathbf{v}} = \mathbf{c}_{\mathbf{u},\mathbf{v}} \mathbf{c}_{\mathbf{v},\mathbf{v}} \, ,
    \end{equation}
    as required. 
    
    To end this example, let us consider the case of the $\mathrm{U}(1)^2$ Chern-Simons theory on the lens space $L(p,1)$ with coupling matrix $\mathbf{C}$ as above. The partition function of this theory is
    \begin{equation}
        \mathcal{Z}_{\mathrm{CS}_{\mathbf{C}}} = \sum_{a_1 = 0}^{p-1} \sum_{a_2 = 0}^{p-1} e^{- 2 i \pi \frac{k}{p} (a_1^2 + a_1 a_2 +a_2^2)} \, ,
    \end{equation} 
    while the associated Reshetikhin-Turaev invariant reads
    \begin{equation}
        \mathcal{RT}_{\mathbf{K}}(\mathbf{L}) = \frac{-i}{\sqrt{3 k^2}} \sum_{u_1 = 0}^{k-1} \sum_{u_2 = 0}^{3k-1} e^{2 i \pi \frac{p}{3k}(3u_1^2 + 3u_1 u_2 + u_2^2)} \, .
    \end{equation}
    It is then an exercise to check that these two quantities fulfill (\ref{result}).
    
    \noindent Let us remark that the following $K$ matrix
    \[ \mathbf{K} = \left( {\begin{array}{*{20}{c}}
    k & 0\\
    0 & 3k
    \end{array}} \right) \, , \]
    also generates the finite group $\mathbb{Z}_{k} \times \mathbb{Z}_{3k}$. However, when $k$ is odd there is no coupling matrix  $\mathbf{C}$ such that  $\mathbf{K} =  \mathbf{C} + \mathbf{C}^\dagger$, and thus no $\mathrm{U}(1)^2$ Chern-Simons theory relying on this $\mathbf{K}$ matrix.  

    4) Let us consider the case of the $\mathrm{U}(1)^3$ Chern-Simons theory with coupling matrix
    \[ \mathbf{C} = \left( {\begin{array}{*{20}{c}}
    k & k & 0 \\
    0 & k & 0 \\
    0 & 0 & k
    \end{array}} \right) \, , \]
    defined on the lens space $L(p,1)$. The associated $\mathbf{K}$ matrix is
    \[ \mathbf{K} = \left( {\begin{array}{*{20}{c}}
    2k & k & 0 \\
    k & 2k & 0 \\
    0 & 0 & 2k
    \end{array}} \right) \, . \]
    It is obvious that this Chern-Simons theory is the direct sum of the one in example 3) with the usual $\mathrm{U}(1)$ Chern-Simons with coupling constant $k$. Accordingly, the partition function of this theory will be the product of the corresponding partition functions, that is to say
    \begin{equation}
        \mathcal{Z}_{\mathrm{CS}} = \left( \sum_{a_1 = 0}^{p-1} \sum_{a_2 = 0}^{p-1} e^{- 2 i \pi \frac{k}{p} (a_1^2 + a_1 a_2 +a_2^2)} \right) \left( \sum_{a_3 = 0}^{p-1} e^{- 2 i \pi \frac{k}{p} a_3^2} \right) \, ,
    \end{equation} 
    Let us recover this factorization in the Reshetikhin-Turaev construction. First, since $det(\mathbf{K}) = 6k^3$, it is quite obvious that $\mathbb{Z}^2/\mathbf{K}\mathbb{Z}^2 = \mathbb{Z}_{k} \times \mathbb{Z}_{2k} \times \mathbb{Z}_{3k}$, with generators
    \[\mathbf{g}_1 = \left( {\begin{array}{*{20}{c}}
    2 \\
    1 \\
    0
    \end{array}} \right) \quad ,\quad \mathbf{g}_{2} = \left( {\begin{array}{*{20}{c}}
    1 \\
    1 \\
    0
    \end{array}} \right)  \quad ,\quad \mathbf{g}_{3} = \left( {\begin{array}{*{20}{c}}
    0 \\
    0 \\
    1
    \end{array}} \right) \, . \]
    Then, with respect to these generators, the linking form $\mathbf{Q}_\mathbf{K}$ associated with $\mathbf{K}$ is
    \[ \mathbf{Q}_\mathbf{K} = \left( {\begin{array}{*{20}{c}}
    \frac{2}{k} & \frac{1}{k} & 0 \\
    \frac{1}{k} & \frac{2}{3k} & 0 \\
    0 & 0 & \frac{1}{2k}
    \end{array}} \right) \, . \]    
    The twist defined by $\mathbf{Q}_\mathbf{K}$ is
    \begin{equation}
    \mathbf{\theta}_{\mathbf{u}} = e^{2 i \pi (\frac{u_1^2}{k} + \frac{u_1 u_2}{k} + \frac{u_2^2}{3k} + \frac{u_3^2}{4k})} \, .
    \end{equation}
    Eventually, we find that the associated Reshetikhin-Turaev invariant is
    \begin{equation}
    \mathcal{RT}_{\mathbf{K}}(\mathbf{L}) = \frac{e^{- \frac{3}{4} i\pi }}{\sqrt{6 k^3}} \sum_{u_1 = 0}^{k-1} \sum_{u_2 = 0}^{3k-1} \sum_{u_3 = 0}^{2k-1} e^{2 i \pi p (\frac{u_1^2}{k} + \frac{u_1 u_2}{k} + \frac{u_2^2}{3k} + \frac{u_3^2}{4k})} \, .
    \end{equation}
    The factorization is then trivial
    \begin{equation}
    \mathcal{RT}_{\mathbf{K}}(\mathbf{L}) = \left( \frac{-i}{\sqrt{3 k^2}} \sum_{u_1 = 0}^{k-1} \sum_{u_2 = 0}^{3k-1} e^{2 i \pi \frac{p}{3k}(3u_1^2 + 3u_1 u_2 + u_2^2)} \right) \left( \frac{e^{- \frac{i \pi }{4} }}{\sqrt{2 k}} \sum_{u_3 = 0}^{2k-1} e^{2 i \pi p \frac{u_3^2}{4k}} \right)\, .
    \end{equation}

	\section{Conclusion}

    In this article we showed that there exists an Reshetikhin-Turaev construction such that the invariant thus generated is equivalent (via some reciprocity procedure) to the partition function of a $\mathrm{U}(1)^n$ Chern-Simons theory. We derived this invariant from a Reshetikhin-Turaev construction whose fundamental ingredient is a twist and not a braiding. This is mainly due to the fact that in this abelian context, the construction only relies on a quadratic form which in turn defines a twist, and then a braiding and an $S$-matrix \cite{GLO18}. Furthermore, the $S$-matrix is not required to be invertible, just like in the case of the $\mathrm{U}(1)$ Chern-Simons theory \cite{MT16}. Let us end this article by pointing out that although to each $\mathrm{U}(1)^n$ Chern-Simons partition function corresponds, through reciprocity, an abelian Reshetikhin-Turaev invariant, the converse is not true. Last but not least, the reciprocity formula can be interpreted as a surgery formula for the $\mathrm{U}(1)^n$ Chern-Simons theory. We know that such a formula also exist for the expectation values of the observables in the $\mathrm{U}(1)$ Chern-Simons theory. Such a surgery formula for observables should also exists for $\mathrm{U}(1)^n$ observables. We know that in the $\mathrm{U}(1)$ Chern-Simons theory zero modes play an important role in the determination of the expectation values  while they are not involved in the determination of the partition function. We expect the same to hold in the $\mathrm{U}^n(1)$ Chern-Simons theory, zero modes having played no role in our study of the partition function of such theory.
    
    As already mentioned, the partition function of the $\mathrm{U}^n(1)$ Chern-Simons theory on an oriented closed smooth  $3$-manifold $M$ only relies on the torsion sector of the first homology group of $M$ and the linking form on it. Consequently, although a surgery presentation of $M$ can be useful to determine these topological data, the only information which remains in the partition function comes from the linking matrix of this surgery presentation. Accordingly, surgery presentations whose linking matrices are stably equivalent \cite{M88}, and thus yield isomorphic finite abelian groups and linking forms, produce the same $\mathrm{U}(1)^n$ Chern-Simons partition function. What is more surprising is that these surgery presentations are always related by a sequence of Kirby moves and Borromeo transformations \cite{M88}.

\end{document}